\documentclass[twocolumn,showpacs,floats,prl,aps,unsortedaddress,superscriptaddress]{revtex4-1}
\usepackage[dvips]{graphicx}
\usepackage{amsfonts}
\usepackage{amsmath}
\usepackage{amssymb}
\usepackage{exscale}
\usepackage{eufrak}
\usepackage{afterpage}
\usepackage{upgreek}
\usepackage{color}	%% FOR COLORED NOTES IN DRAFT. COMMENT THIS LINE FOR FINAL VERSION 
\usepackage{braket}	%% For nice BRAKET notation
\usepackage{multirow}	%% For columns spanning multiple rows in Tables
\usepackage{enumitem}

\usepackage[normalem]{ulem}    %% To use \uline for underlining, 
\begin{document}

\title{High-density two-dimensional electron system induced by oxygen vacancies in ZnO.}
%%%%%%%%

\author{T.~C.~R\"odel}
\thanks{T.C.~R\"odel and J.Dai contributed equally to this work}
\affiliation{CSNSM, Univ. Paris-Sud, CNRS/IN2P3, Universit\'e Paris-Saclay, 
			91405 Orsay Cedex, France}
\affiliation{Synchrotron SOLEIL, L'Orme des Merisiers, Saint-Aubin-BP48, 
			91192 Gif-sur-Yvette, France}
\affiliation{Laboratory for Photovoltaics, Physics and Material Science Research Unit, 
			University of Luxembourg, L-4422 Belvaux, Luxembourg}

\author{J.~Dai}
\thanks{T.C.~R\"odel and J.Dai contributed equally to this work}
\affiliation{CSNSM, Univ. Paris-Sud, CNRS/IN2P3, Universit\'e Paris-Saclay, 
			91405 Orsay Cedex, France}

\author{F.~Fortuna}
\affiliation{CSNSM, Univ. Paris-Sud, CNRS/IN2P3, Universit\'e Paris-Saclay, 
			91405 Orsay Cedex, France}
			
\author{E.~Frantzeskakis}
\affiliation{CSNSM, Univ. Paris-Sud, CNRS/IN2P3, Universit\'e Paris-Saclay, 
			91405 Orsay Cedex, France}

\author{P.~Le~F\`evre}
\affiliation{Synchrotron SOLEIL, L'Orme des Merisiers, Saint-Aubin-BP48, 
			91192 Gif-sur-Yvette, France}

\author{F.~Bertran}
\affiliation{Synchrotron SOLEIL, L'Orme des Merisiers, Saint-Aubin-BP48, 
			91192 Gif-sur-Yvette, France}

\author{M.~Kobayashi}
\affiliation{Photon Factory, Institute of Materials Structure Science,
			High Energy Accelerator Research Organization (KEK), 
			1-1 Oho, Tsukuba 305-0801, Japan}

\author{R.~Yukawa}
\affiliation{Photon Factory, Institute of Materials Structure Science,
			High Energy Accelerator Research Organization (KEK), 
			1-1 Oho, Tsukuba 305-0801, Japan}

\author{T.~Mitsuhashi}
\affiliation{Photon Factory, Institute of Materials Structure Science,
			High Energy Accelerator Research Organization (KEK), 
			1-1 Oho, Tsukuba 305-0801, Japan}
			
\author{M.~Kitamura}
\affiliation{Photon Factory, Institute of Materials Structure Science,
			High Energy Accelerator Research Organization (KEK), 
			1-1 Oho, Tsukuba 305-0801, Japan}

\author{K.~Horiba}
\affiliation{Photon Factory, Institute of Materials Structure Science,
			High Energy Accelerator Research Organization (KEK), 
			1-1 Oho, Tsukuba 305-0801, Japan}

\author{H.~Kumigashira}
\affiliation{Photon Factory, Institute of Materials Structure Science,
			High Energy Accelerator Research Organization (KEK), 
			1-1 Oho, Tsukuba 305-0801, Japan}

\author{A.~F.~Santander-Syro}
\email{andres.santander@csnsm.in2p3.fr}
\affiliation{CSNSM, Univ. Paris-Sud, CNRS/IN2P3, Universit\'e Paris-Saclay, 
			91405 Orsay Cedex, France}

\date{\today}

% 79.60.-i Photoemission and photoelectron spectra
%%%%%%%%%%%%%%%

%%%%%%%%%%%%%%%
\begin{abstract}
%%%
We realize a two-dimensional electron system (2DES) in ZnO 
by simply depositing pure aluminum on its surface in ultra-high vacuum,
and characterize its electronic structure using angle-resolved photoemission spectroscopy. 
The aluminum oxidizes into alumina by creating oxygen vacancies 
that dope the bulk conduction band of ZnO and confine the electrons near its surface. 
The electron density of the 2DES is up to two orders of magnitude
higher than those obtained in ZnO heterostructures. 
The 2DES shows two $s$-type subbands, that we compare to the $d$-like 2DESs in titanates, 
with clear signatures of many-body interactions that we analyze 
through a self-consistent extraction of the system self-energy and a modeling as a coupling 
of a 2D Fermi liquid with a Debye distribution of phonons.
%%%
\end{abstract}
%%%%%%%%%%%%%%%
%
\maketitle

%%%%%%%%%%%%%%%%%%%%%%%%%%%%%%%%%%%%%%%%%%%%%%%%%%%%%%%%%%%%%%%%%%%%%%%%%%%%%%%%%%%%%%%
%%%%%%%%%%% INTRODUCTION
%%%%%%%%%%%%%%%%%%%%%%%%%%%%%%%%%%%%%%%%%%%%%%%%%%%%%%%%%%%%%%%%%%%%%%%%%%%%%%%%%%%%%%%
ZnO is a transparent, easy to fabricate, oxide semiconductor with a direct band gap $E_g = 3.3$~eV. 
Its many uses include window layers in photovoltaic devices, 
varistors for voltage surge protection, UV absorbers, gas sensors, 
and catalytic devices~\cite{ZnO-book,Book-ZnO-Wiley}.
ZnO is also a candidate for novel applications, such as transparent 
field effect transistors, UV laser diodes, memristors, or high-temperature/high-field 
electronics~\cite{ZnO-book,Book-ZnO-Wiley,Lorenz2016,Liu2015,Bakin2007,Ozgur2005}.
In fact, ZnO can be seen as a link between the classical group-IV or III-V semiconductors, 
e.g. Si or GaAs, and transition metal oxides (TMOs), such as SrTiO$_3$.
Due to their valence $d$-orbitals, the latter show a rich variety of collective electronic phenomena, 
like magnetism or high-$T_c$ superconductivity~\cite{Tokura2000,Dagotto2005}.
%%%%%%
Moreover, the controlled fabrication of a two-dimensional electron system (2DES) in ZnO 
can result in extremely high electron mobilities, 
even competing with the ones of GaAs-based heterostructures, 
and showing the quantum hall effects~\cite{Tsukazaki2007,Tsukazaki2010}. 
%%%

2DES in TMOs have also improved in mobility over the last decade, 
but what makes them really unique is the control
of superconductivity, magnetism and spin-orbit coupling 
by varying the electron density using a gate voltage~\cite{Ohtomo2004,Thiel2006,Reyren2007,
Ueno2008,Brinkman2007,Caviglia2008,Caviglia2010,Joshua2013,
Chen2013,Chen2015,Ngo2015,Stornaiuolo2016}.  
Additionally, for many insulating TMOs, 
recent works demonstrated that oxygen vacancies near the surface 
provide a simple and efficient mechanism to produce a 2DES, 
with electron densities as high as
$n_{2D} \sim 3 \times 10^{14}$~cm$^{-2}$~\cite{Santander-Syro2011,Meevasana2011,
Santander-Syro2012,King2012,Bareille2014,Roedel2014,Walker2014,Roedel2015,
Roedel2016,Roedel2017,Frantzeskakis2017}, and showing magnetic states
linked to the presence of such vacancies~\cite{Taniuchi2016}. 
%%%

%%%%%%%%%%%%%%%%%
%% HERE WE SHOW
%%%%%%%%%%%%%%%%%
Here we show, using angle-resolved photoemission spectroscopy (ARPES), 
that the simple evaporation in ultra-high vacuum (UHV) 
of an atomic layer of pure aluminum on ZnO creates a 2DES with electron densities 
up to two orders of magnitude higher than in previous studies. 
We demonstrate that the 2DES results from oxidation of the Al layer and
concomitant doping with oxygen vacancies of the underlying ZnO surface.
The 2DES is composed of two subbands with different effective masses, 
as the mass of the inner band is wholly renormalized 
due to the energetic proximity of its band bottom with a phonon frequency, 
whereas the outer band, dispersing deeper in energy, shows only a kink 
due to the electron-phonon interaction. 
We thoroughly investigate the electron-phonon coupling by a self-consistent extraction 
of the electron self-energy.
We deduce an Eliashberg coupling function wholly compatible 
with a 2D Debye-like distribution of phonons and a mass enhancement parameter $\lambda=0.3$. 
%%%%%%%%%%%%%%%%%%

%%%
Previous photoemission experiments on ZnO~\cite{Powell1972,Gopel1982,
Ozawa2005,Ozawa2009,Ozawa2010,Piper2010,Deinert2015}
showed that hydrogenation of its polar or non-polar surfaces, 
for instance through chemisorption of hydrogen, methanol or water, 
induces a downward band-bending and the formation of a 2DES
with a moderate electron density $n_{2D} \leq 2 \times 10^{13}$~cm$^{-2}$,
showing only one broad shallow subband below the Fermi level ($E_F$)~\cite{Piper2010}.
%%%%
More recently, several ARPES studies 
focused on the many-body phenomena of electron-phonon coupling in oxides, 
demonstrating that at \emph{low carrier densities} 
the 2DES in TiO$_2$, SrTiO$_3$ and also ZnO 
are composed of polarons~\cite{Moser2013,Wang2016,Yukawa2016}. 
Due to a non-adiabatic electron-phonon coupling, 
the polaronic regime changes to a Fermi liquid behavior with increasing electron densities, 
as electronic screening of the polar lattice becomes more efficient~\cite{Verdi2017}. 
However, the Fermi liquid regime in ZnO
% \textcolor{red}{
% expected at carrier densities well above an estimated critical value
% $n_{2D}^{c} \approx 4 \times 10^{12}$~cm$^{-2}$~\cite{Verdi2017},}
has not been studied yet, 
as previous doping methods of the surface were insufficient 
to achieve high electron densities.
Attaining large carrier densities for a 2DES in ZnO 
is also appealing for applications in high-power transparent electronics.
%%%%%%%%%%%%

%%%%%%%%%%%%%%%%%%%%%%%%%%%%%%%%%%%%%%%%%%%%%%%%%%%%%%%%%%%%%%%%%%%%%%%%%%%%%%%%%%%%%%%
%%%%%%%%%%% RESULTS
%%%%%%%%%%%%%%%%%%%%%%%%%%%%%%%%%%%%%%%%%%%%%%%%%%%%%%%%%%%%%%%%%%%%%%%%%%%%%%%%%%%%%%%
%%%%%%%%%%%%%%%%%%%%%%%%%%%%%%%%
\begin{figure}[tb]
        \includegraphics[clip, width=0.48\textwidth]{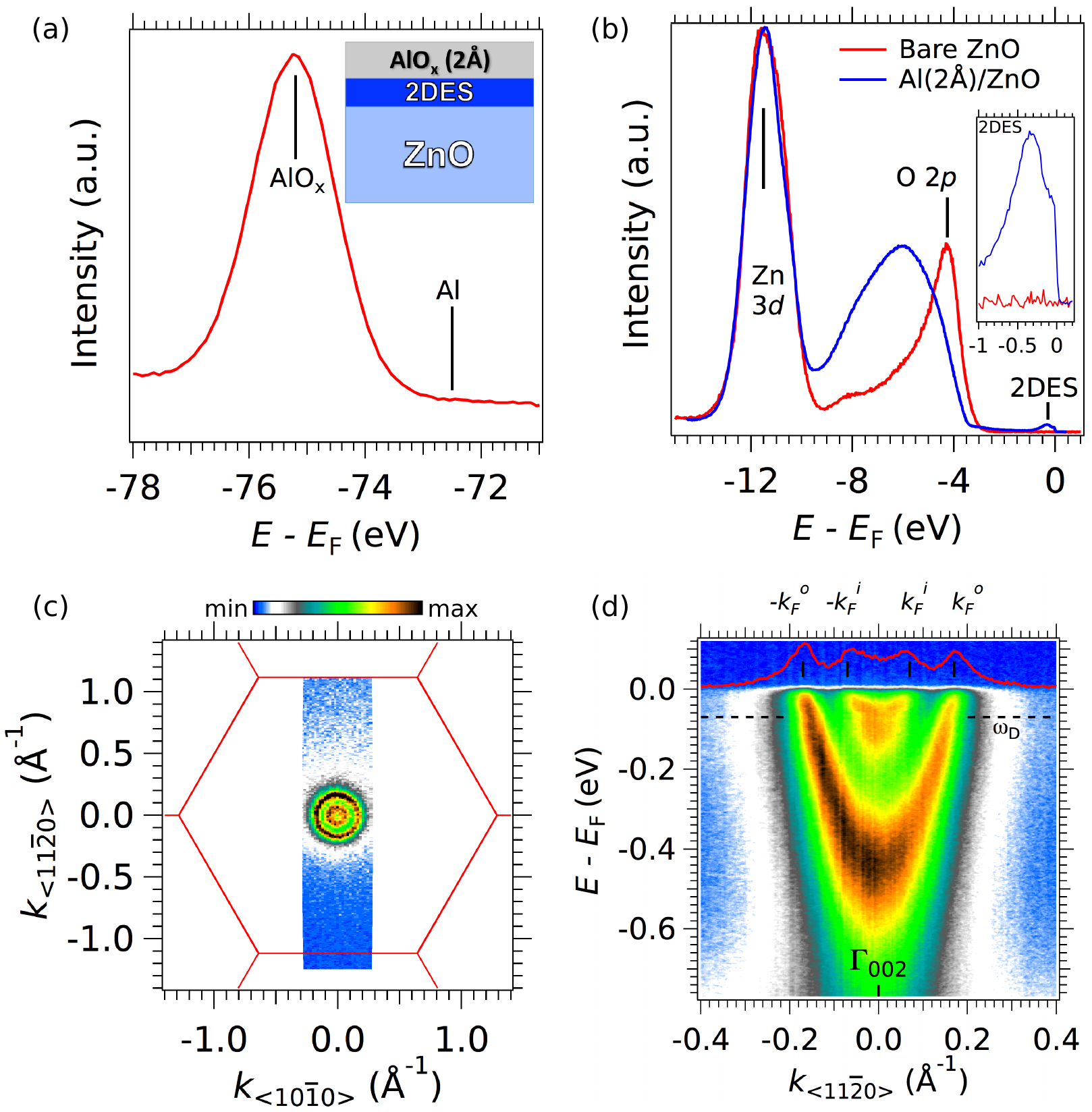}
    \caption{\label{fig:ZnO-FS} {(Color online)
        	 (a) Al $2p$ core level, measured right after deposition on ZnO.
        	 Its binding energy corresponds to \emph{completely oxidized} aluminum (AlO$_x$), 
        	 while no peak of pure aluminum is observed. 
        	 The inset shows the AlO$_x$/ZnO interface and the 2DES obtained 
        	 after deposition of Al on ZnO. 
        	 (b)~Comparison of the valence bands and near-$E_F$ spectra of bare ZnO
        	 and Al-capped ZnO (red and blue curves, respectively). The surface redox reaction
        	 after Al deposition modifies the O $2p$ valence band, and
        	 produces an intense quasi-particle peak at $E_F$, corresponding to the 2DES.
        	 As no Fermi level was detected on the spectrum at the bare ZnO surface, 
        	 its binding energies were calibrated with respect to the Zn 3$d$ peak 
        	 of the AlO$_x$/ZnO interface.
        	 (c)~ARPES Fermi surface map of the Al(2~\AA)/ZnO interface
        	 in the O-terminated $[000\bar{1}]$ plane, measured at $h\nu = 88$~eV 
        	 with linear-horizontal light polarization. 
        	 Red lines indicate the edges of the in-plane Brillouin zone.
			 (d)~Energy-momentum ARPES intensity map around the bulk $\Gamma_{002}$ point
			 along the in-plane $k_{<11\bar{2}0>}$ direction, measured at $h\nu = 25$~eV 
        	 with linear-horizontal light polarization. The red curve is the MDC
			 over $E_F \pm 5$~meV. 
             The black vertical bars show the Fermi momenta $k_F^i$ and $k_F^o$ 
             of the inner and outer subbands. 
             The Fermi liquid is coupled to phonons with a characteristic Debye 
             energy $\omega_D$ shown by the horizontal, dashed black line.
             Photoemission data in this and all other figures of this paper were measured
        	 at $T=7$~K.
             %%%
        }
      }
\end{figure}
%%%%%%%%%%%%%%%%%%%%%%%%%%%%%%%%
%%

%%%%%%%%%%%%%%%%%%%%%%%%%%%%%%%%%%%%%%%%%%%%%%
%%%%%%%%% Effects Al-capping: Ovacs and 2DES
%%%%%%%%%%%%%%%%%%%%%%%%%%%%%%%%%%%%%%%%%%%%%%
We now discuss our main findings.
Henceforth, we will focus on data measured at the O-terminated ZnO$(000\bar{1})$ surface.
As shown in the Supplementary Material, similar results
are obtained at the ZnO$(0001)$ (zinc-terminated) interface,
although the resulting 2DES has a slightly smaller electron density.
%%%%%%%%%%%%%
Furthermore, to recall that we deposited pure Al (not aluminum oxide) on the ZnO surface, 
we note the resulting AlO$_x$ capping layer simply as ``Al",
specifying in parenthesis the evaporated thickness.
Additional details on the crystallographic nomenclature, surface preparation, aluminum evaporation, 
and ARPES measurements are provided in the Supplementary Material.
%%%

%%%
The creation of a 2DES using Al deposition is identical to the procedure described 
in Ref.~\cite{Roedel2016}. It is worth noting that, for previously reported 2DES in oxides, 
the intense synchrotron beam can create oxygen vacancies due to desorption induced by electronic
transitions~\cite{Walker2015}. This process, based on the photo-excitation  of core levels, 
is different in titanates and ZnO~\cite{Tanaka2004}. 
Thus, our results demonstrate that the creation of 2DES in oxides using Al 
is a much more general mechanism, enabling furthermore ARPES studies 
independent from the relaxation mechanism  of photo-excited core levels.
%%%

%%%
Fig.~\ref{fig:ZnO-FS}(a) shows that the Al-$2p$ core-level peak 
at the Al(2~\AA)/ZnO interface corresponds to oxidized aluminum, 
whose binding energy ($E-E_F = -75$~eV) is very different from the one
of metallic aluminum ($-72.5$~eV)~\cite{Roedel2016}.
Fig.~\ref{fig:ZnO-FS}(b) compares the valence-band 
of the bare, stoichiometric ZnO$(000\bar{1})$ surface (red curve) and
of the Al(2~\AA)/ZnO interface (blue curve).
We observe that, contrary to oxygen-deficient surfaces or interfaces 
of TMOs~\cite{Roedel2016,Backes2016}, there are no measurable 
states corresponding to localized electrons (i.e., deep donors)
in the band gap of oxygen-deficient ZnO. 
The absence of such states in ZnO emphasizes the simpler character of a 2DES 
based on $s$-valence electrons, compared to the $d$-valence electrons in TMOs. 
On the other hand, 
the binding energy and shape of the O $2p$ valence band are dramatically changed, 
possibly because the O-$2p$ valence band of the oxidized Al layer is at a binding energy 
of $\approx 6$~eV.
Moreover, as detailed in the inset of Fig.~\ref{fig:ZnO-FS}(b),
the Al(2~\AA)/ZnO interface shows a clear 
quasi-particle peak at $E_F$, not present at the bare surface.
%%%

The contribution of oxygen vacancies to $n$-type conductivity in bulk ZnO 
has been a controversial issue~\cite{Janotti2005,Janotti2007,Janotti2009,Lany2010,Kim2009}.  
The photoemission signatures observed here after Al deposition, 
namely an oxidized Al core level and the appearance of a 2DES at $E_F$,
are identical to the ones reported in other oxides~\cite{Roedel2016},
indicating that the mechanisms underlying the 2DES formation are similar. 
Future theoretical works should explore in detail
the energetics and specific role of oxygen vacancies near the \emph{surface} of ZnO.
%%%%%

%%%%%%%%%%%%%%%%%%%%%%%%%%%%%%%%%%%%%%%%%%
%%%%%%%%% Description 2DES band structure
%%%%%%%%%%%%%%%%%%%%%%%%%%%%%%%%%%%%%%%%%%
We now characterize the electronic structure of the 2DES
at the Al(2~\AA)/ZnO$(000\bar{1})$ (oxygen-terminated) interface.
Fig.~\ref{fig:ZnO-FS}(c) shows the in-plane Fermi surface map measured by ARPES. 
There are two metallic states
forming in-plane circular Fermi sheets around $\Gamma$,
that correspond to confined states of ZnO's conduction band 
--which is formed by orbitals of $s$-character.
Fig.~\ref{fig:ZnO-FS}(d) presents the energy-momentum dispersion map 
of the two states forming the above concentric Fermi circles,
henceforth called outer ($o$) and inner ($i$) subbands.
They were measured around the bulk $\Gamma_{002}$ point
along the in-plane $k_{<11\bar{2}0>}$ direction. 
Such 2DES with two subbands in ZnO had not been observed before, 
as electron densities were not large enough in previous 
studies~\cite{Gopel1982,Ozawa2005,Ozawa2009,Ozawa2010,Piper2010,Deinert2015,Yukawa2016}. 
%%%
Additional data presented in the Supplementary Material
demonstrates that the in-plane periodicity of the electronic structure 
corresponds to the one of an unreconstructed surface,
and that the two subbands form cylindrical, non-dispersive Fermi surfaces 
along the $(0001)$ direction perpendicular to the interface, 
confirming their 2D character.
%%%%%

\begin{figure}[tb]
        \includegraphics[clip, width=0.48\textwidth]{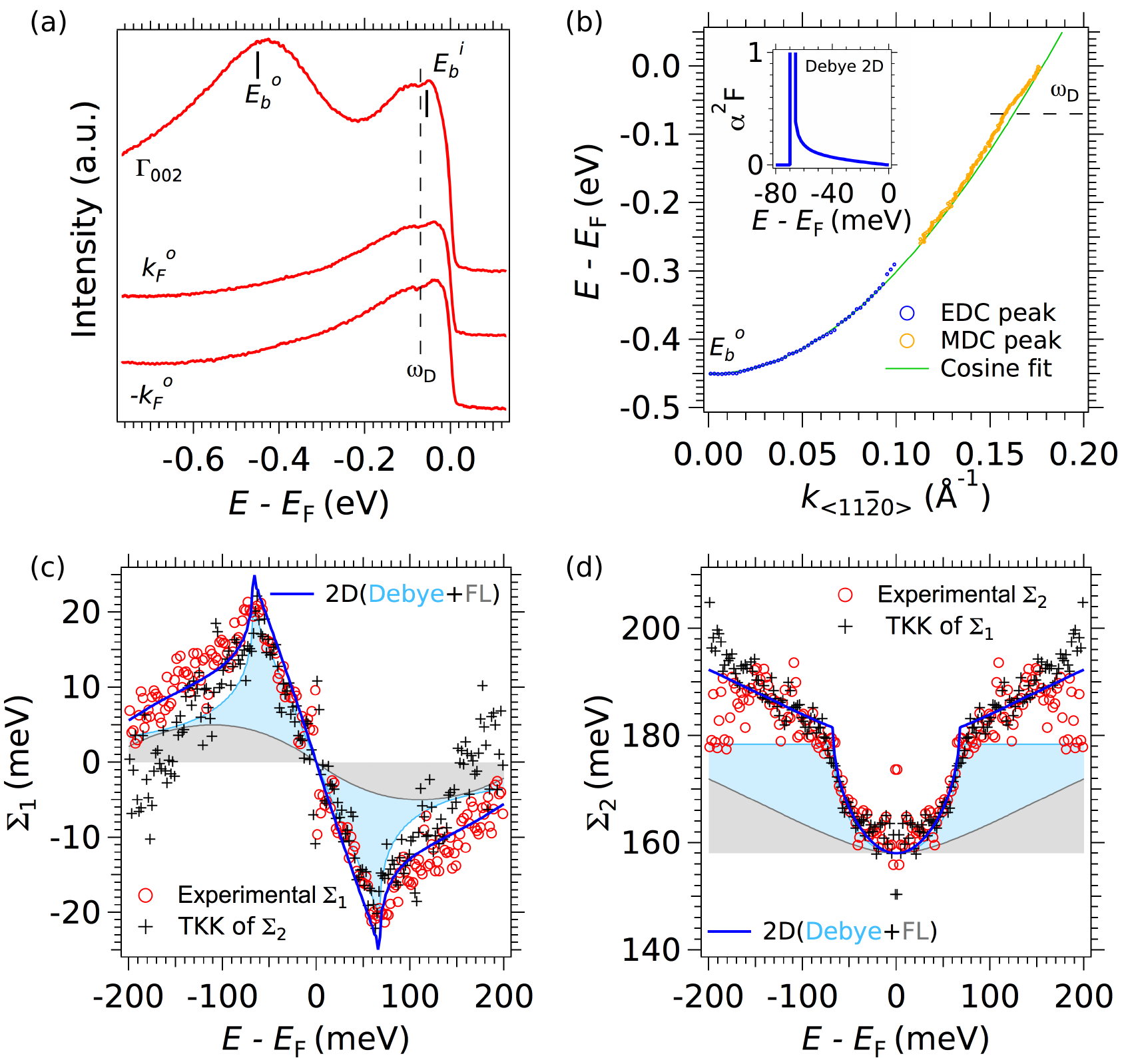}
    \caption{\label{fig:ZnO-SelfEn} {(Color online)
        	 (a)~EDCs of the ARPES dispersion map, Fig.~\ref{fig:ZnO-FS}(b), 
			 respectively over $\Gamma \pm 0.05$~\AA$^{-1}$ (upper curve),
			 and $\pm k_F^o\pm 0.05$~\AA$^{-1}$ (mid and lower curve).
			 The peaks near $E_F$ for both the inner and outer subbands 
			 show a peak-dip-hump structure, with the dip at an energy 
			 $\omega_D = -70$~meV. 
        	 (b)~Maxima of the EDC (blue circles) and MDC (orange circles) peaks
        	 for the outer subband of the 2DES the Al(2~\AA)/ZnO$(000\bar{1})$ interface. 
        	 Only data for the right branch ($k > 0$) are shown.
        	 The continuous green curve is a cosine fit to the data
        	 representing the non-interacting electron dispersion of this subband.
        	 (c,~d)~Experimental real and imaginary parts of the electron self-energy (red circles),
        	 and their Kramers-Kronig transforms (black crosses), 
        	 for the right branch of the outer subband. 
        	 The dark blue curves are simultaneous fits to $\Sigma_1$ and $\Sigma_2$ 
        	 using a 2D Fermi-liquid~$+$~Debye model.
        	 The 2D Fermi liquid and 2D Debye components of the fit are shown
        	 by the filled light blue and gray curves.
        	 Data were symmetrized with respect to $E_F$, as required
        	 by Kramers-Kronig.
        	 Similar results are obtained by an analysis of the left branch
        	 of the outer subband (see Supplementary Material).
        	 The inset in (b) shows the Eliashberg coupling function 
        	 resulting from the used 2D Debye model.
        }
      }
\end{figure}

%%%
The subbands' Fermi momenta, determined from the maxima of the 
momentum distribution curve (MDC) integrated over $E_F \pm 5$~meV, 
red curve on top of Fig.~\ref{fig:ZnO-FS}(d), 
are $k_F^o = (0.17 \pm 0.005)$~\AA$^{-1}$ and $k_F^i = (0.07 \pm 0.005)$~\AA$^{-1}$.
Their band bottoms, extracted from the maxima of the energy distribution curve (EDC)
over $\Gamma \pm 0.05$~\AA$^{-1}$ and the dispersion of the EDC peaks,
Figs.~\ref{fig:ZnO-SelfEn}(a,~b), 
are located at binding energies $E_b^o = (450 \pm 5)$~meV 
and $E_b^i = (55 \pm 5)$~meV. 
Due to the light and isotropic band mass of the $s$-type electrons forming the 2DES, 
the subband splitting in ZnO is $\approx 3$ times larger 
than in titanates~\cite{Roedel2017}. 
The thickness of the 2DES can be estimated from the subbands' 
binding energies and energy separation by assuming a triangular-wedge quantum well,
yielding $21$~\AA~(or $4$ unit cells) along $c$
(see Supplementary Material for details).
%%%%%

%%%%%
From the area enclosed by the in-plane Fermi circles ($A_F$), 
the density of electrons in the 2DES is 
$n_{2D} = A_F/(2 \pi^2) = (5.4 \pm 0.3) \times 10^{13}$~cm$^{-2}$,
or about $0.14$ electrons per hexagonal unit cell in the $(000\bar{1})$ plane.
Such electron density is far larger than the critical value, 
estimated at $3.8 \times 10^{12}$~cm$^{-2}$, at which the crossover from a polaronic 
to a Fermi liquid regime for electron-phonon coupling occurs~\cite{Verdi2017}.
Additionally, the effective masses around $\Gamma$ of the outer and inner subbands, 
determined from their Fermi momenta and band bottoms 
using free-electron parabola approximations, 
are respectively $m_o^{\star} = (0.25 \pm 0.02) m_e$ 
and $m_i^{\star} = (0.34 \pm 0.08) m_e$,
where $m_e$ is the free-electron mass.
The mass of the outer subband agrees well with the conduction-band mass 
along the $(000\bar{1})$ plane calculated for bulk stoichiometric ZnO
or determined from infrared reflectivity and cyclotron resonance
experiments on lightly-doped ZnO~\cite{ZnO-book,Book-ZnO-Wiley}.
As the confinement of non-interacting electrons in a quantum well 
should result in subbands with the same effective mass, 
we will focus on analyzing the renormalization of inner band in the following paragraphs.
%%%%

%%%%%%%%%%%%%%%%%%%%%%%%%%%%
%% Analysis of e-ph coupling
%%%%%%%%%%%%%%%%%%%%%%%%%%%%
%%%%%%%%%%%%%%%%%%%%%%%%%%%%%%%%
%%%%%%%%%%%%%%%%%%%%%%%%%%%%%%%%

%%%%%%%%%%%%%%%%%%%%%%%%%%%%%%%%
\begin{figure}[tb]
        \includegraphics[clip, width=0.48\textwidth]{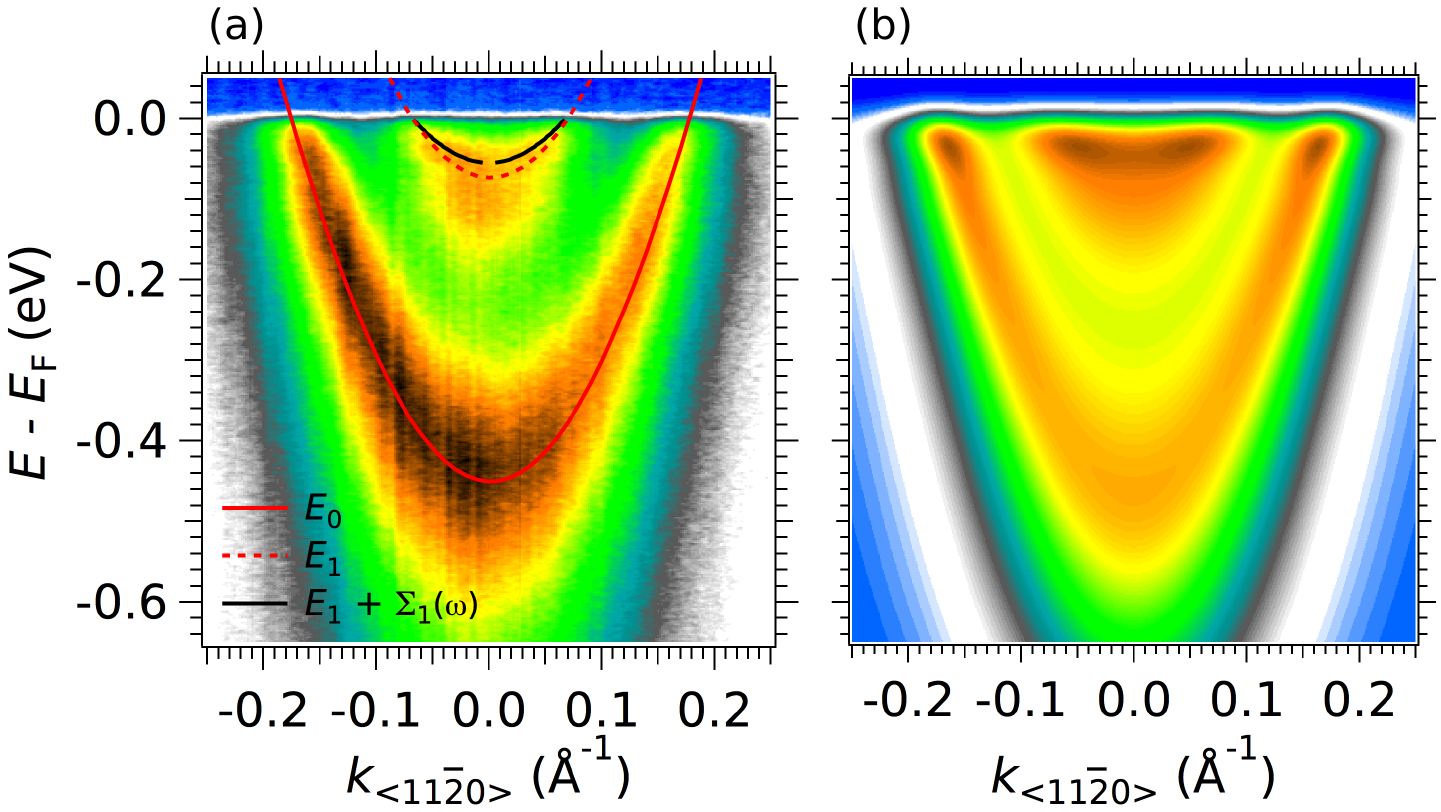}
    \caption{\label{fig:ZnO-SimSelfEn} {(Color online)
        	 (a)~Zoom over the energy-momentum ARPES map of Fig.~\ref{fig:ZnO-FS}(b).
        	 The continuous red curve is the cosine fit to the bare outer subband.
        	 The dashed red curve is the same fit up-shifted by $377$~meV, 
        	 until it matches the Fermi momenta of the inner subband,
        	 representing thus the bare inner subband.
        	 The black curve is the \emph{renormalized} inner subband, obtained
        	 by adding the experimental $\Sigma_1$ extracted from the \emph{outer} subband,
        	 Fig.\ref{fig:ZnO-SelfEn}(c), to the bare \emph{inner} subband.
        	 (b) Simulation of the ARPES data, using the self-energy of the 
        	 2D Fermi-liquid~$+$~Debye model fitted to the data, 
        	 plus a constant electronic scattering rate of approximately $160$~meV,
        	 corresponding to $\Sigma_2$ at $E_F$ from Fig.~\ref{fig:ZnO-SelfEn}(d). 
        	 The calculated spectral function was 
        	 multiplied by the Fermi-Dirac distribution and then convoluted with a 
        	 Gaussian resolution function of $\rm{FWHM} = 20$~meV. 
        }
      }
\end{figure}
%%%%%%%%%%%%%%%%%%%%%%%%%%%%%%%%

%%%%%
In fact, as seen from Figs.~\ref{fig:ZnO-FS}(d) and~\ref{fig:ZnO-SelfEn}(a),
the band bottom of the \emph{inner} subband presents a complex structure, 
with a peak-dip-hump clearly seen in the EDC around $\Gamma$. 
Likewise, as shown in Figs.~\ref{fig:ZnO-FS}(d) and~\ref{fig:ZnO-SelfEn}(a,~b),
the \emph{outer} subband shows a \emph{kink} in its dispersion 
at approximately the same binding energy ($\omega_D = 70$~meV)
of the dip observed in the inner band, together with a peak-dip-hump for the EDCs 
around its Fermi momenta. 
As will be shown shortly, all these features result from electron-phonon coupling.

%%%%
The quantification of electron-phonon interaction 
is possible through the analysis of the energy-dependent real ($\Sigma_1$) 
and imaginary ($\Sigma_2$) parts of the electron self-energy. 
These can be inferred from the spectral function of the many-electron
system, directly measured by ARPES~\cite{Hufner-Book}.
Thus, we will extract and model the self energy of the outer band, and then use the 
results to renormalize the inner band, which is difficult to fit 
due to the the peak-dip-hump structure.  
%%%

%%%
Fig.~\ref{fig:ZnO-SelfEn}(b) shows the dispersion of the spectral function peak
for the outer subband, extracted from the maxima of the EDCs (blue circles)
and MDCs (orange circles). The continuous green line is a cosine fit to the data
representing the bare (i.e., non-interacting) electron dispersion of this subband.
The energy difference between the MDC peak and the bare band 
gives the real part of the electron self-energy, 
and is plotted in Fig.~\ref{fig:ZnO-SelfEn}(c), red circles. 
The pronounced \emph{peak} in $\Sigma_1$ at $E-E_F \approx -70$~meV 
corresponds to the kink in the experimental dispersion.
Likewise, the energy dependence of the MDCs line-widths gives the imaginary part 
of the electron self energy (or electronic scattering rate), 
and is shown in Fig.~\ref{fig:ZnO-SelfEn}(d), red circles.
Here, one observes a rapid increase of the scattering rate from $E_F$ down to the
binding energy at which the real part of the self-energy peaks, 
followed by a less rapid but steady increase.
To check the consistency of the self-energy extracted from our data, 
we compute the Kramers-Kronig transformation (TKK) of the experimental $\Sigma_2(E)$ 
[respectively $\Sigma_1(E)$], black crosses in Fig.~\ref{fig:ZnO-SelfEn}(c) 
[respectively Fig.~\ref{fig:ZnO-SelfEn}(d)]. We observe an excellent agreement
between $\Sigma_1(E)$ and $\textrm{TKK}\lbrace \Sigma_2(E) \rbrace$
[respectively between $\Sigma_2(E)$ and $\textrm{TKK}\lbrace \Sigma_1(E) \rbrace$],
ensuring that our analysis and choice of bare dispersion respect causality.
%%%%

%%
The simultaneous occurence of a pronounced peak in $\Sigma_1$ 
and an abrupt change in slope in $\Sigma_2$ at about the same energy $\omega_D$,
as observed in Figs.~\ref{fig:ZnO-SelfEn}(c,~d), 
are typical landmarks of the interaction between the electron liquid
and some collective modes of the solid (e.g. phonons) 
having a characteristic energy $\omega_D$~\cite{Hufner-Book}.
Thus, we fit the experimental complex self-energy 
with a model of a Fermi liquid with Debye electron-phonon coupling, 
both in 2D~\cite{Book-ElectronLiquids2005,Kostur1993},
as shown by the continuous blue curves in Figs.~\ref{fig:ZnO-SelfEn}(c,~d). 
The fit gives a Debye frequency of $68 \pm 2$~meV, in excellent agreement
with our data and the phonon energies (up to about $580$~cm$^{-1}$, 
or $\approx 70$~meV) measured by other techniques~\cite{Ozgur2005,ZnO-book}, 
and a dimensionless coupling constant $\lambda = 0.3 \pm 0.05$,
The isotropic Fermi liquid of the fit is characterized by a carrier density
of $(6.7 \pm 0.4) \times 10^{13}$~cm$^{-2}$, close to the experimental value.
The electron-phonon, or Eliashberg, coupling function $\alpha^2 F(\omega)$ 
resulting from the used 2D Debye model is shown in the inset of Fig.\ref{fig:ZnO-SelfEn}(b).
%%%%%%%%%%
%% Leave Debye 3D for SupMat
%%%%%%%%%% 
We checked that a fit with a 3D Fermi liquid~$+$~Debye 
model~\cite{Book-ElectronLiquids2005,Engelsberg1963} 
yields a larger phonon cutoff energy, 
of the order of $85$~meV, and an overall poor agreement with the experimental self-energy. 
The details of the models and a comparison of the obtained fits 
are given in the Supplementary Material.
%%%%

%%%%%
We now turn to the inner subband. To model it, we rigidly shift the bare outer band in energy 
and then renormalize it using the previously deduced self-energy. 
As shown by the red curve in Fig.~\ref{fig:ZnO-SimSelfEn}(a), a shift of $377$~meV 
fits the experimental Fermi momenta, matching the required conservation 
of the 2D electron density.  
The resulting renormalized inner band, black curve in Fig.~\ref{fig:ZnO-SimSelfEn}(a), 
compares excellently with the experimental inner band.
%%%

%%%
To cross-check the above analysis of the ARPES data,
we simulated the whole 2DES spectral function using the self-energy of the 
2D Fermi liquid~$+$~Debye model fitted to the data. The resulting ARPES map,
Fig.~\ref{fig:ZnO-SimSelfEn}(b), compares well with the data. 
Thus, the \emph{entire} electronic structure of the 2DES at the 
Al(2~\AA)/ZnO$(000\bar{1})$ surface can be understood from doping 
of the bulk conduction band by oxygen vacancies, 
electron confinement due to band-bending induced by those vacancies, 
and coupling of the ensuing subbands with a Debye-like distribution of phonons.
%%%%%%%%%%%%%

%%%%%
Note that, in the present case of a high carrier density, 
the coupling constant $\lambda$ gives directly the electron mass renormalization $m^{\star}$
due to electron-phonon interaction, namely $m^{\star}/m_0 = 1 + \lambda$~\cite{Giustino2016},
where $m_0$ is the non-interacting band mass.
%%%
Using parabolic approximations (i.e., energy-independent band masses) 
for the subbands' dispersions, we can assume that the bottom of the outer band, 
located well below the phonon energies, gives the non-interacting band mass, 
while the inner band, located just above the Debye energy,
gives the electron mass fully renormalized by coupling to phonons.
This yields a coupling constant 
$\lambda \approx 1- m_i^{\star}/m_o^{\star} = 0.36 \pm 0.3$,
subject to large errors, but in overall agreement with the more accurate value obtained 
above from the fit to the whole energy-dependent complex self-energy.
%%%%

%%%
More generally, in insulating dielectric oxides, the electron-phonon coupling 
can significantly depend on the electron density,
due to different screening mechanism of the oscillating ions. 
At low densities (i.e. band fillings smaller or comparable to the phonon cutoff frequency),
screening based on dielectric polarization results in large, spatially delocalized, polarons. 
At high densities the increased electronic screening 
of the ionic lattice vibrations result in a Fermi liquid regime 
with weaker electron phonon coupling~\cite{Verdi2017}.
%%%
Those two regimes were recently characterized by ARPES in 
anatase-TiO$_2$(001)~\cite{Moser2013} and SrTiO$_3$~\cite{Wang2016}.
%%%
Note furthermore that the electron-phonon coupling constant $\lambda = 0.3$ obtained here
is significantly smaller than the coupling constant observed in the Fermi liquid regime
of anatase-TiO$_2$ and SrTiO$_3$ 
($\lambda_{FL}^{\rm{TiO}_2} \approx \lambda_{FL}^{\rm{SrTiO}_3} = 0.7$)~\cite{Moser2013,
Meevasana2010,Wang2016}.
This suggests that, in the high carrier density regime, the electronic screening 
for electron-phonon coupling is more efficient for the $s$ electrons of the 2DESs in ZnO 
than for the $d$ electrons of the 2DESs in TMOs.
%%%%%

%%%%%%%%%%%%%%%%%%%%%%%%%%%%%%%%%%%%%%%%%%%%%%%%%%%%%%%%%%%%%%%%%%%%%%%%%%%%%%%%%%%%%%%
%%%%%%%%%%% CONCLUSIONS
%%%%%%%%%%%%%%%%%%%%%%%%%%%%%%%%%%%%%%%%%%%%%%%%%%%%%%%%%%%%%%%%%%%%%%%%%%%%%%%%%%%%%%%
Notably, our self-consistent Kramers-Kronig analysis of the self energy in ZnO, 
and the deduction of the electron-phonon coupling parameter using a Debye model, 
is different from previous approaches used in other oxides, like SrTiO$_3$~\cite{King2014},
where $\lambda$ was inferred from the slope of $\Sigma_1$ at $E_F$ 
(i.e., the renormalization of quasiparticle mass),
or TiO$_2$ anatase, where it was estimated by modeling the self-energy 
to reproduce the data~\cite{Moser2013}. Note that the coupling parameters deduced from the 
renormalization of quasiparticle mass, velocity and spectral weight in ARPES data 
are in general subject to large errors, as mentioned before, and distinct from the true microscopic 
coupling parameter~\cite{Veenstra2010}. 
%%%

As a whole, our results highlight the universal character of the approach 
based on surface redox reactions to create 2DESs in functional oxides~\cite{Roedel2016}, 
unveil similarities and differences between $s$- and $d$-orbital type 2DES,
and add new ingredients to the rich many-body physics 
displayed by confined electronic states in ZnO. 
Our observations suggest that oxygen vacancies can contribute to electron-doping 
near the surface of ZnO, motivating further experimental and theoretical studies
on the formation and role of vacancies at surfaces/interfaces of this important
transparent semiconductor oxide.
Moreover, the realization of a highly doped 2DES in ZnO 
opens a new realm of possibilities, such as high-power applications 
using a transparent oxide semiconductor that presents many advantages with respect
to standard Sn-doped In$_2$O$_3$ (ITO): 
ZnO is more abundant, cheaper,  easier to fabricate and process, 
non toxic, and when doped it can attain mobilities comparable 
to those of ITO~\cite{ZnO-book,Book-ZnO-Wiley,Lorenz2016}. 

%%%%%%%%%%%%%%%%%%%%%%%%%%%%%%%%%%%%%%%%%%%%%%%%%%%%%%%%%%%%%%%%%%%%%%%%%%%%%%%%%%%%%%%
%%%%%%%%%%%%%%%%%%%%%%%%%%%%%%%%%%%%%%%%%%%%%%%%%%%%%%%%%%%%%%%%%%%%%%%%%%%%%%%%%%%%%%%
%%%%%%%%%%%%%%%%%%%%%%%%%%%%%%%%%%%%%%%%%%%%%%%%%%%%%%%%%%%%%%%%%%%%%%%%%%%%%%%%%%%%%%%
\acknowledgments 
% We thank \textcolor{red}{XYZ} for discussions.
Work at CSNSM was supported by public grants from the French National Research Agency (ANR), 
project LACUNES No ANR-13-BS04-0006-01, 
and the ``Laboratoire d'Excellence Physique Atomes Lumi\`ere Mati\`ere'' 
(LabEx PALM projects ELECTROX and 2DEG2USE) overseen by the ANR as part of the 
``Investissements d'Avenir'' program (reference: ANR-10-LABX-0039).
Work at KEK-PF was supported by Grants-in-Aid for Scientific Research 
(Nos. 16H02115 and 16KK0107) from the Japan Society for the Promotion of Science (JSPS).
Experiments at KEK-PF were performed under the approval of the 
Program Advisory Committee (Proposals 2016G621 and 2015S2005) 
at the Institute of Materials Structure Science at KEK.
T.~C.~R. acknowledges funding from the RTRA--Triangle de la Physique (project PEGASOS).
A.F.S.-S. thanks support from the Institut Universitaire de France.
%%%%%%%%%%%%%%%%%%%%%%%%%%%%%%%%%%%%%%%%%%%%%%%%%%%%%%%%%%%%%%%%%%%%%%%%%%%%%%%%%%%%%%%
%%%%%%%%%%%%%%%%%%%%%%%%%%%%%%%%%%%%%%%%%%%%%%%%%%%%%%%%%%%%%%%%%%%%%%%%%%%%%%%%%%%%%%%
%%%%%%%%%%%%%%%%%%%%%%%%%%%%%%%%%%%%%%%%%%%%%%%%%%%%%%%%%%%%%%%%%%%%%%%%%%%%%%%%%%%%%%%

%%%%%%%%%%%%%%%%%%%%%%%%%%%%
%%%% SUPPLEMENTARY MATERIAL
%%%%%%%%%%%%%%%%%%%%%%%%%%%%
\section*{SUPPLEMENTARY MATERIAL}

%%%%%%%%%%%%%%%%%%%%%%%%%%%%%
\subsection*{ZnO structure and notation}
%%%%%%%%%%%%%%%%%%%%%%%%%%%%%%%%
%%%
ZnO crystallizes in the hexagonal wurtzite structure, with the oxygen anions
forming a tetrahedron around the Zn cation. The lattice constants are
$a = 3.25$~\AA~and $c = 5.2$~\AA.
All through this paper, we use the Miller-Bravais 4-index, or $(hkil)$, 
convention for hexagonal systems, which makes permutation symmetries apparent,
where $(hkl)$ are the regular Miller indices for an hexagonal lattice,
and the third (redundant) index is defined as $i = -(h+k)$.
Thus, we note $[hkil]$ the crystallographic directions in real space,  
$(hkil)$ the planes orthogonal to those directions, 
and $\langle hkil \rangle$ the corresponding directions in reciprocal space.
%%%%

%%%%%%%%%%%%%%%%%%%%%%%%%%%%%
\subsection*{Surface preparation and Aluminum deposition}
%%%%%%%%%%%%%%%%%%%%%%%%%%%%%%%%
%%%
%%%%%%%%%%%%%%%%%%%%%%%%%%%%%%%%
\begin{figure}[b]
        \includegraphics[clip, width=0.48\textwidth]{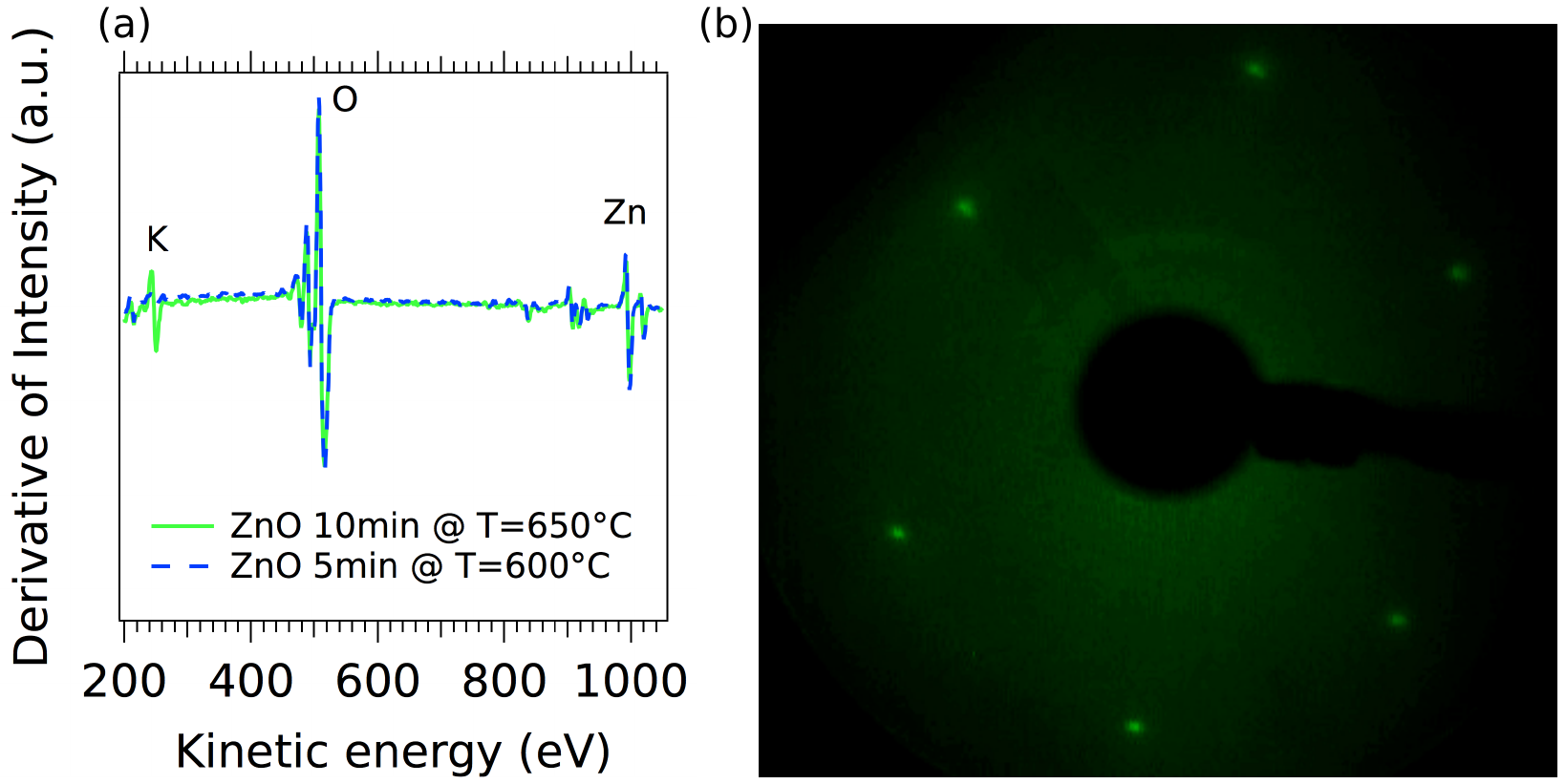}
    \caption{\label{fig:ZnO-LEED-Auger} {%(Color online)
        	 (a)~Auger spectra of the oxygen-terminated ZnO$(000\bar{1})$
        	 surface. Depending on the preparation protocol, 
        	 potassium diffuses from the bulk to the surface as evidenced 
        	 by the peak at a kinetic energy of $E_{kin} = 250$~eV. 
        	 (b)~LEED image of oxygen-terminated ZnO$(000\bar{1})$ surface 
        	 after sputtering and annealing at $T=650^{\circ}$C.
        }
      }
\end{figure}
%%%%%%%%%%%%%%%%%%%%%%%%%%%%%%%%
%%%
There are two possible terminations of the polar $(0001)$ ZnO surface: 
Zn, or $(0001)$ termination, and O, or $(000\bar{1})$ termination~\cite{Mariano1963}. 
Commercially available single crystals (SurfaceNet GmbH) 
with the two terminations at opposing faces were used in our experiments. 
The surfaces of ZnO were prepared based on the work 
of Dulub and coworkers~\cite{Dulub2002}. 
The existence of potassium impurities in single crystals of ZnO is possible 
as potassium hydroxide is an educt in the hydrothermal synthesis of single crystals. 
Long annealing at elevated temperatures resulted in the migration of potassium impurities 
from the bulk to the surface as evidenced by the Auger spectra in 
Fig.~\ref{fig:ZnO-LEED-Auger}(a). 

We followed two procedures to reduce the presence of potassium 
and obtain an atomically clean and cristalline surface:

\begin{enumerate}[topsep=0pt, itemsep=0pt, parsep=0pt, partopsep=5pt]
	\item Ar+ sputtering at $1$~kV for 10~minutes,
	\item short annealing for 5 min at $T \approx 600$C to $700^{\circ}$C in UHV;
\end{enumerate}

or alternatively:

\begin{enumerate}[topsep=0pt, itemsep=0pt, parsep=0pt, partopsep=5pt]
	\item Ar+ sputtering at $1$~kV for 10~minutes at $T \approx 600-700^{\circ}$C,
	\item stop annealing approximately 5~minutes after sputtering.
\end{enumerate}

These procedures resulted in LEED images similar to the one shown 
in Fig.~\ref{fig:ZnO-LEED-Auger}(b). 

Oxygen vacancies were then created on the UHV clean and cristalline ZnO surfaces
by the deposition of 2~\AA~of aluminum at sample temperatures $T \approx 50-100^\circ$C.
The complete details on the Al deposition are described elsewhere~\cite{Roedel2016}. 
%%%%%%%%%%%%%%%%%%%%%%%%%%%%%

%%%%%%%%%%%%%%%%%%%%%%%%%%%%%
\subsection*{Photoemission measurements}
%%%%%%%%%%%%%%%%%%%%%%%%%%%%%
ARPES experiments were performed at the CASSIOPEE beamline of Synchrotron SOLEIL (France)
and at beamline 2A of KEK-Photon Factory (KEK-PF, Japan)
using hemispherical electron analyzers with vertical and horizontal slits, respectively. 
Pristine sample surfaces and oxygen-vacancy doping by Al-capping were obtained 
by \textit{in situ} surface preparation as described in the previous section.
The sample temperature during measurements was 7 K (SOLEIL) or 20 K (KEK-PF), 
without observing any $T$-dependence between these two temperature values. 
The typical angular and energy resolutions were 0.25$^{\circ}$ and 15 meV, 
while the mean diameter of the incident photon beam was of the order of 
50 $\mu$m (SOLEIL) and 100 $\mu$m (KEK-PF). 
We used variable energy and polarization of the incident photons. 
A systematic variation of the photon energy revealed no changes 
in the energy-momentum dispersion (see next section), 
a feature that is characteristic of 2D-like band structures. 
During the time window of our measurements the pressure was in the range 
of 10$^{\textmd{-}11}$ mbar and no evolution or degradation of the spectra
was observed.
%%%%%

%%%%%%%%%%%%%%%%%%%%%%%%%%%%%
\subsection*{Electronic structure of 2DES: in-plane periodicity and out-of-plane confinement}
%%%%%%%%%%%%%%%%%%%%%%%%%%%%%
%%%
%%%%%%%%%%%%%%%%%%%%%%%%%%%%%%%%
\begin{figure*}[tb]
        \includegraphics[clip, width=\textwidth]{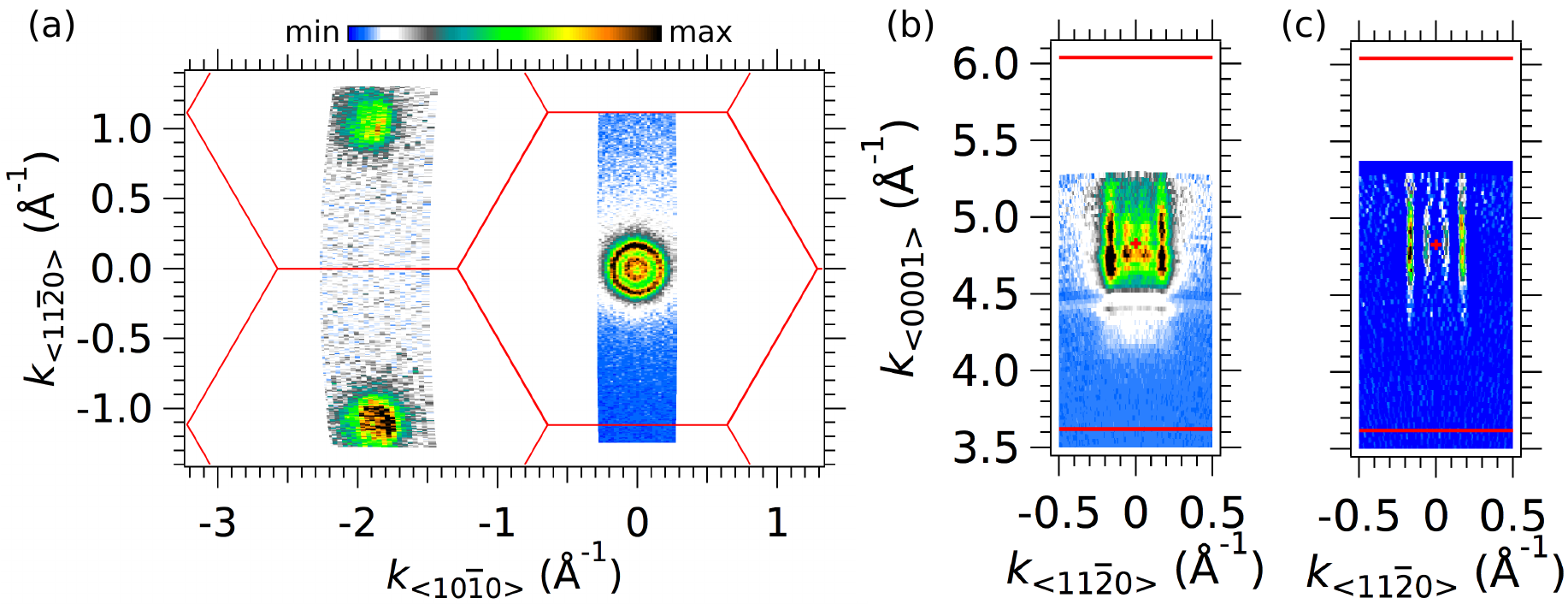}
    \caption{\label{fig:ZnO-FS-extended} {%(Color online)
        	 (a)~ARPES Fermi surface map of the Al(2~\AA)/ZnO interface
        	 in the $[000\bar{1}]$ plane, extended over several Brillouin zones. 
        	 Data was measured at $h\nu = 88$~eV 
        	 with linear-horizontal light polarization.
        	 Red lines indicate the edges of the in-plane Brillouin zone.
        	 (b)~Fermi surface map along the $(0001)$ (out-of-plane) direction,
        	 in the $k_{<000\bar{1}>} - k_{<1\bar{1}20>}$ plane,
        	 measured by changing the photon energy between 
        	 $h\nu = 40$~eV and $h\nu = 105$~eV in steps of 1 eV. 
        	 A free-electron final-state model~\cite{Hufner-Book} 
        	 with inner potential of $V_0 = 5$~eV was assumed to calculate 
        	 the $k_{<000\bar{1}>}$ values.
        	 Red lines indicate the edges of the bulk out-of-plane Brillouin zone, 
        	 and the red cross shows the location of a bulk $\Gamma$ point.
        	 There is no measurable spectral weight at the Fermi level between
        	 $k_{<000\bar{1}>} \approx 2.7$~\AA$^{-1}$ and 
        	 $k_{<000\bar{1}>} \approx 4.3$~\AA$^{-1}$.
        	 (C)~Second derivative (negative values) of the data in~(b). 
        	 The cylindrical shape of the two out-of-plane Fermi surfaces is clear.
        	 All measurements were conducted at $T = 7$~K.
        }
      }
\end{figure*}
%%%%%%%%%%%%%%%%%%%%%%%%%%%%%%%%
%%%
Fig.~\ref{fig:ZnO-FS-extended}(a) shows the in-plane Fermi surface
measured by ARPES, extended over three neighboring Brillouin zones
of the unreconstructed ZnO$(000\bar{1})$ surface. 
The two concentric circular Fermi sheets described in the main text
are systematically observed around each of the $\Gamma$ points in these
Brillouin zones, demonstrating that the electronic structure
has the periodicity expected from an unreconstructed surface.
Figs.~\ref{fig:ZnO-FS-extended}(b,~c) furthermore show that such two states 
form cylindrical, non-dispersive Fermi surfaces 
in the $k_{\langle 11\bar{2}0 \rangle} - k_{\langle 0001 \rangle}$ plane,
\emph{i.e.} along the $(0001)$ direction perpendicular to the interface, 
confirming their 2D character.
Note that the Fermi surface in the $k_{\langle 11\bar{2}0 \rangle} - k_{\langle 0001 \rangle}$
plane would be circular in the case of a 3D state, 
as the effective mass of the $s$-electrons is isotropic.
%%%%%

%%%%%%%%%%%%%%%%%%%%%%%%%%%%%
\subsection*{Confinement potential and extension of the 2DES at the AlO$_x$/ZnO interface}
%%%%%%%%%%%%%%%%%%%%%%%%%%%%%
%%%
The characteristics of the confinement potential, 
assumed for simplicity as triangular-wedge shaped,
can be readily extracted from the bottom energies
of the outer and inner subbands~\cite{Santander-Syro2011}. 
As the lowest edge of ZnO's conduction-band 
is mainly $s$-like~\cite{ZnO-book}, the out-of-plane effective masses, 
which enter into the computation of the quantum well eigen-energies, should be identical
to the in-plane masses directly determined from our ARPES data. 
Thus, using the effective mass around $\Gamma$
of the \emph{outer} subband, which is non-renormalized by electron-phonon interaction,
and the energy difference of $380$~meV between the outer and inner subbands, 
we find that the 2DEG realized in our experiments corresponds to electrons
confined by a field of about $F \approx 280$~MV$/$m in a well of depth $V_0 \approx -0.93$~eV.
The geometrical depth of the quantum well is then $d = V_0/eF \approx 33$~\AA.
The thickness of the 2DEG can also be estimated from the average position 
of the inner subband's wave-function, corresponding to the electrons in the quantum well
farthest away from the surface. From the solutions to the Schr\"odinger equation 
in the above potential wedge, this yields approximately $21$~\AA,
or about 4 unit cells along $c$, in agreement with the value inferred from the 
quantum-well geometrical depth.  
%%%%%

%%%%%%%%%%%%%%%%%%%%%%%%%%%%%
\subsection*{Debye model for electron-phonon coupling}
%%%%%%%%%%%%%%%%%%%%%%%%%%%%%
%%%%%%%%%%%%%%%%%%%
\subsubsection*{Eliashberg formalism}
%%%%%%%%%%%%%%%%%%%
The theoretical tool to deal with the Hamiltonian including the electron-phonon
interaction is the Eliashberg theory, at the center of which 
is the Eliashberg coupling function, 
$\alpha^2 F(\omega)$~\cite{frohlich1950theory,echenique2004decay,hofmann2009electron}. 
This function can be interpreted as the phonon density of states (at energy $\omega$)
weighted by the electron-phonon coupling matrix element.

The electron-phonon mass enhancement parameter $\lambda$, 
can be calculated from $\alpha^2 F(\omega)$ and understood 
as the dimensionless coupling strength:

\begin{equation}
	\label{lambda}
	% \lambda(\varepsilon_i, \bold{k})=2\int_0^{\omega_{max}}
	\lambda = 2\int_0^{\omega_{max}}
	\frac{\alpha^2 F(\omega)}{\omega} d\omega.
\end{equation}

Here, $\omega_{max}$ is the maximum phonon energy, 
and usually takes the value of Debye energy $\omega_D$. 
The factor 2 appears as both the absorption and emission processes are counted in. 

In the limit $T\rightarrow 0$K that we will use, the electron-phonon self-energy
can also be calculated from the Eliashberg coupling function as:

\begin{equation}
	\label{zero_temp}
	\Sigma_2(\omega,\bold{k}; T=0)=\pi \int_0^{min(\omega,\omega_{max})} 
	\alpha^2 F(\omega')d\omega'.
\end{equation}

In principle, once the dispersion relation of the scattering phonon is given, 
the Eliashberg coupling function and the electron-phonon coupling strength can be calculated 
by applying some assumptions and approximations. 
In simple models, such as the Einstein model and the Debye model, 
analytical results can be derived. 
%%%

%%%%%%%%%%%%%%%%%
\subsubsection*{Self-energy for the 2D and 3D Debye models}
%%%%%%%%%%%%%%%%%
%%%
%%%%%%%%%%%%%%
%% Debye 2D
%%%%%%%%%%%%%%
In the 2D Debye model, used in the main text to analyze our data, 
the Eliashberg coupling function $\alpha^2 F(\omega)$ 
can be analytically calculated~\cite{Kostur1993}:

%%%
\begin{equation}
	\label{2d_debye}
	\alpha^2 F(\omega)=
	\begin{cases}
		\frac{\lambda}{\pi} \frac{\omega}{(\omega_D^2-\omega^2)^{1/2}}, 
		& \lvert\omega\rvert < \omega_D \\
        0, & \lvert\omega\rvert > \omega_D 
	\end{cases}
\end{equation}   
%%%

In turn, from Eq.~\ref{zero_temp} and Eq.~\ref{2d_debye}, 
$\Sigma_2$ can be also calculated:

\begin{equation}
	\Sigma_2^{2D}(\omega)=
	\begin{cases}
		\lambda \omega_D(1-\sqrt{1-(\frac{\omega}{\omega_D})^2}),  & \lvert\omega\rvert<\omega_D \\
        \lambda \omega_D, &\lvert\omega\rvert>\omega_D. 
	\end{cases}
\end{equation}

Here, $\lambda$ is the mass enhancement parameter, 
also given by the negative slope of $\Sigma_1$ at $E-E_F=0$. 
%%%

%%%
The real part of the self-energy for the 2D Debye model ($\Sigma_1^{2D}$)
does not have an analytical expression, but can be readily calculated 
from the Hilbert transform of $\Sigma_2^{2D}$.

%%%%%%%%%%%%%%
%% Debye 3D
%%%%%%%%%%%%%%
The 3D Debye model has simple analytical forms 
for both $\Sigma_1$ and $\Sigma_2$ \cite{Engelsberg1963}. 
For instance, for $\Sigma_2$ it is: 

%%%
\begin{equation}\label{3D_debye}
	\Sigma_2^{3D}(\omega)=
	\begin{cases}
		\frac{\pi}{3}\lambda \omega_D(\frac{\omega}{\omega_D})^3,  & \lvert\omega\rvert<\omega_D \\
        \frac{\pi}{3}\lambda \omega_D, & \lvert\omega\rvert>\omega_D 
	\end{cases}
\end{equation}
%%%

Thus, using either the 2D or 2D Debye models, one can extract accurate values of
the Debye energy $\omega_D$ and the dimensionless coupling strength $\lambda$ 
from fits to the whole energy dependent complex self-energy.
%%

%%%
%%%%%%%%%%%%%%%%%%%%%%%%%%%%%%%%
\begin{figure}[tb]
        \includegraphics[clip, width=0.48\textwidth]{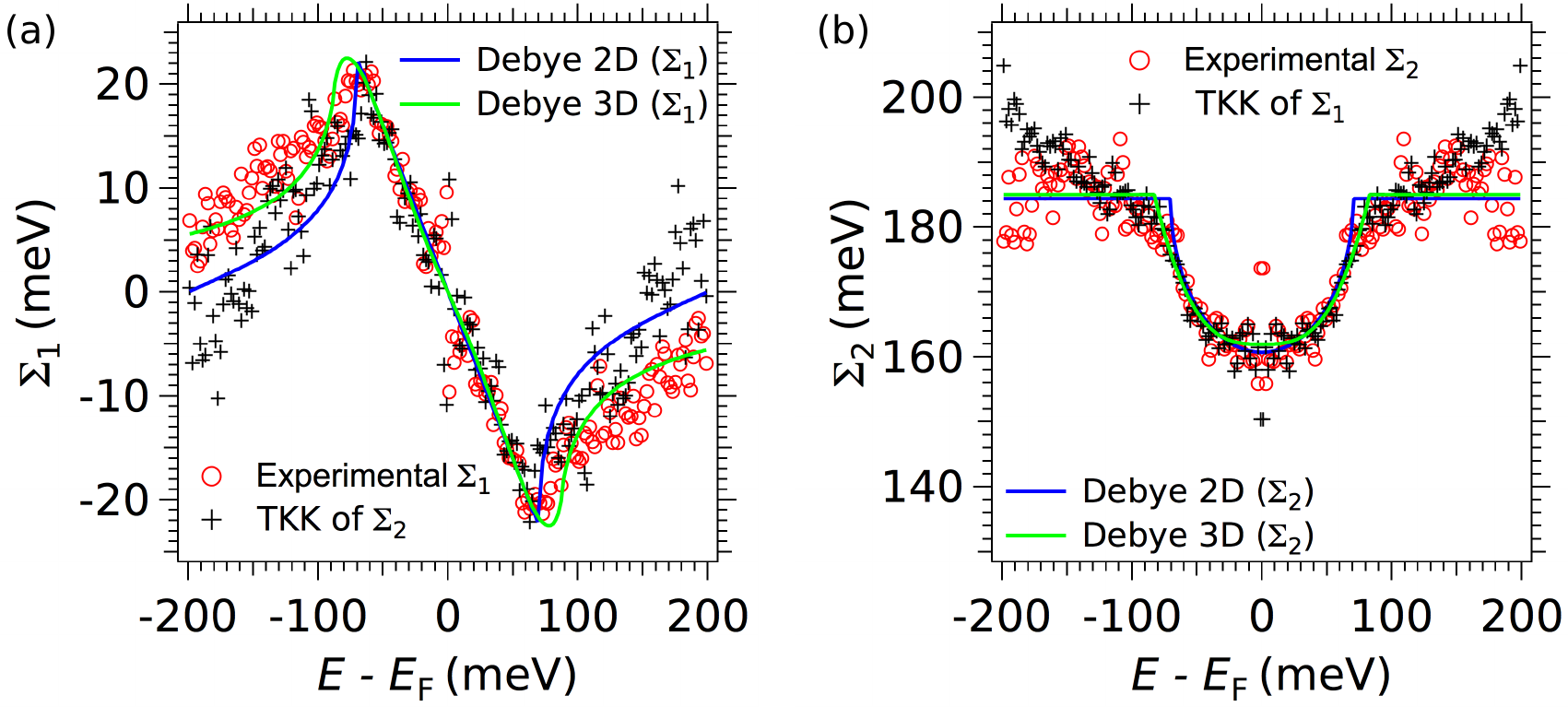}
    \caption{\label{fig:ZnO-2DEG-Debye2D-3D} {%(Color online)
             Comparison of fits to the experimental self-energy using Debye models 
             in 2D (blue lines, $\omega_D = 70$~meV, $\lambda = 0.34$) 
             and 3D (green lines, $\omega_D = 85$~meV, $\lambda = 0.27$).
        	 (a)~Real part of the self-energy.
        	 (b)~Imaginary part of the self-energy.
        	 For simplicity, we have omitted in the fits the Fermi-liquid part of the self-energy,
        	 which only contributes to small corrections above $\omega_D$.
        }
      }
\end{figure}
%%%%%%%%%%%%%%%%%%%%%%%%%%%%%%%%
%%%

Figure~\ref{fig:ZnO-2DEG-Debye2D-3D} compares 
the fitting results with 3D and 2D Debye models for the outer subband 
(right branch, as in the main text) of the 2DES in ZnO.
%%%%
From this figure, it appears that the 3D Debye model fits better
$\Sigma_1$ when $\lvert \omega \rvert > \omega_D$, while the 2D Debye model 
shows a sharper inflection at $\lvert \omega \rvert = \omega_D$. 
As for the fitting of $\Sigma_2$, the two models give similar results, 
except that the 3D model tends to give a larger $\omega_D$ 
and a flatter bottom of $\Sigma_2$, in less good agreement with experiments. 
%%%

%%%%%%%%%%%%%%%%%%%%%%%%%%%%%
\subsection*{Fermi-liquid self-energy in 2D and 3D}
%%%%%%%%%%%%%%%%%%%%%%%%%%%%%
%%%
%%%%%%%%%%%%%%%%%%%%%%%%%%%%%%%%
\begin{figure}[tb]
        \includegraphics[clip, width=0.48\textwidth]{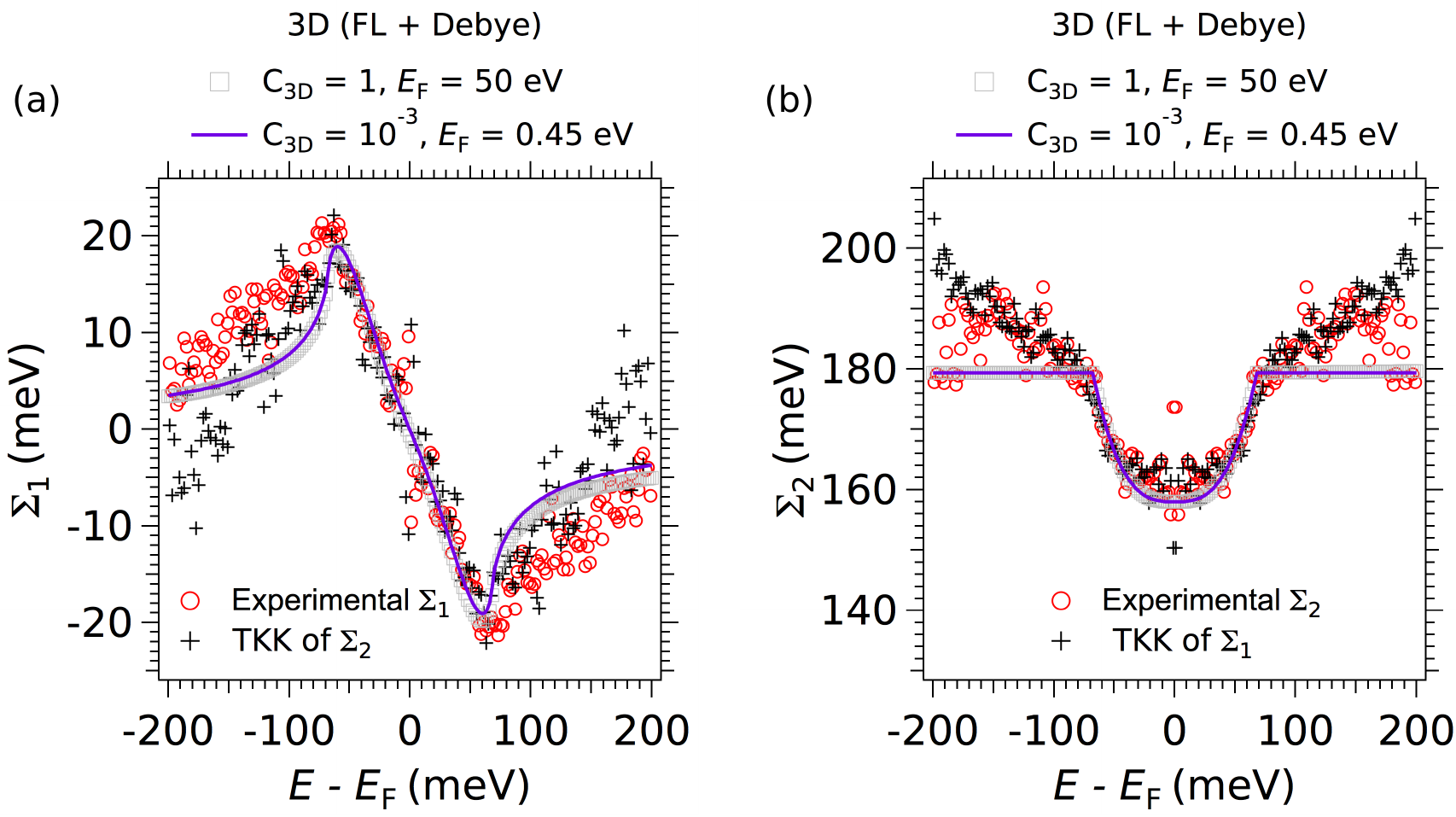}
    \caption{\label{fig:ZnO-2DEG-FL-Debye-2D-3D} {%(Color online)
             Fits to the experimental self-energy using 3D models for the 
             Fermi-liquid and Debye electron-phonon coupling.
        	 (a)~Real part of the self-energy.
        	 (b)~Imaginary part of the self-energy.
        	 In both fits, the same parameters for the Debye frequency ($\omega_D = 68$~meV)
        	 and electron-phonon coupling constant $\lambda = 0.3$ were used.
        	 The best fit parameters for the Fermi-liquid are unphysical: 
        	 either the constant $C_{3D}$ is very small 
        	 or the band filling $\epsilon_F$ is exaggeratedly large.  
        	 All other fits with physically reasonable parameters 
        	 completely fail to capture the value and energy-dependence
        	 of the self energy, especially above $\omega_D$.
        }
      }
\end{figure}
%%%%%%%%%%%%%%%%%%%%%%%%%%%%%%%%
%%%
At temperatures much smaller than the bottom $\epsilon_F$ of the conduction band
(bare electron mass $m_e$), the self-energy of a 2D electron liquid 
can be written as~\cite{Book-ElectronLiquids2005}:
%%%
\begin{equation}\label{2D_FL}
	\Sigma_2^{\rm FL-2D}(\varepsilon)=
		C_{2D}\frac{\varepsilon^2}{4 \pi \epsilon_F}
		\ln |\frac{4 \epsilon_F}{\varepsilon}|, 
\end{equation}
%%%
where $C_{2D}$ is a constant, $\epsilon_F = 2 \pi n_{2D} \times \hbar^2/(2 m_e)$,
and $n_{2D}$ is the density of electrons.
%%%

Similarly, the self-energy of a 3D electron liquid is~\cite{Book-ElectronLiquids2005}:
%%%
\begin{equation}\label{3D_FL}
	\Sigma_2^{\rm FL-3D}(\varepsilon)=
		C_{3D}\frac{\pi}{8 \epsilon_F}
		\frac{\varepsilon^2 + (\pi k_B T)^2}{1 + e^{-\beta \varepsilon}}, 
\end{equation}
%%%
where $C_{3D}$ is a constant, and $\epsilon_F$ is the bottom of the conduction band
for the electron system.
%%%

In both cases, the real part of the self-energy can be calculated 
from the Hilbert transform of $\Sigma_2$.

Fits to the experimental self-energy using 2D models for the Fermi liquid and the
Debye electron-phonon coupling where presented in the main text.
Figure~\ref{fig:ZnO-2DEG-FL-Debye-2D-3D} shows fits with 3D models 
of a Fermi-liquid~$+$~Debye self-energy for the outer subband 
(right branch, as in the main text) of the 2DES in ZnO.
%%%%
It is clear that the 2D model used in the main text provides a much better fit. 
Moreover, the use of a 3D Fermi liquid model yields ``reasonable'' (albeit still poor) 
fits of the experimental data only for unphysical values of the model parameters,
such as a vanishingly small constant $C_{3D} \ll 1$ 
(for metallic systems, it should be close to 1~\cite{Book-ElectronLiquids2005})
or an exceedingly large $\epsilon_F \approx 50$~eV.  
%%%

%%%%%%%%%%%%%%%%%%%%%%%%%%%%%
\subsection*{Experimental self-energy for the left branch of the outer subband}
%%%%%%%%%%%%%%%%%%%%%%%%%%%%%
%%%
%%%%%%%%%%%%%%%%%%%%%%%%%%%%%%%%
\begin{figure*}[bht]
        \includegraphics[clip, width=\textwidth]{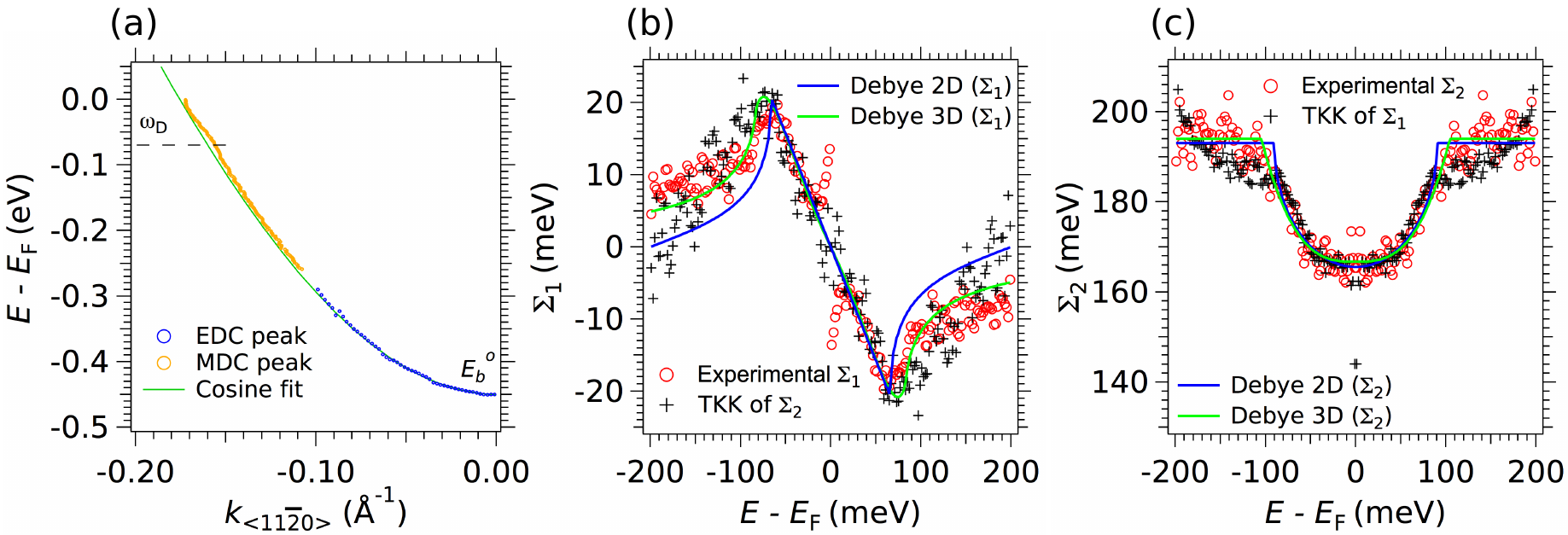}
    \caption{\label{fig:ZnO-2DEG-SigmaLEFT} {%(Color online)
        	 (a)~Maxima of the EDC (blue circles) and MDC (red circles) peaks
        	 for the \emph{left} branch ($k < 0$) of the outer subband 
        	 of the 2DES the Al(2~\AA)/ZnO$(000\bar{1})$ interface. 
        	 The continuous green curve is the same cosine fit to the data
        	 used for the right branch (main text),
        	 representing the non-interacting electron dispersion of this subband.
        	 (b,~c)~Experimental real and imaginary parts of the electron self-energy (red circles),
        	 and their Kramers-Kronig transforms (black crosses), 
        	 for the \emph{left} branch of the outer subband.
        	 The blue and green curves are simultaneous fits to $\Sigma_1$ and $\Sigma_2$ 
        	 using, respectively, 2D and 3D Debye models for the self-energy.
        	 For simplicity, we have omitted the Fermi-liquid part of the self-energy,
        	 which contributes only to small corrections above $\omega_D$.
        }
      }
\end{figure*}
%%%%%%%%%%%%%%%%%%%%%%%%%%%%%%%%
%%%
%%%%%%%%%%%%%%%%%%%%%%%%%%%%%%%%
\begin{table}[bht]
	\begin{tabular}{c|c|c|c|c}
		\hline
		\hline
     	                  & $L_1$ & $L_2$ & $R_1$ & $R_2$ \\
		\hline
		$\omega_D/\rm{eV}(2D)$ & 0.066 & 0.099  &  0.070 & 0.070 \\
		\hline
		$\omega_D/\rm{eV}(3D)$ & 0.084 & 0.10 & 0.088 & 0.083 \\
		\hline
		$\lambda(2D)$     & 0.33 & 0.276  & 0.34  & 0.34 \\
		\hline
		$\lambda(3D)$    & 0. 27 & 0.25 & 0.28 & 0.27 \\
		\hline
	\end{tabular}
	\caption{
			Comparison of the Debye frequencies and electron-phonon coupling constants 
			extracted from fits to the experimental self-energy 
			using the 2D and 3D Debye models.
			The Fermi-liquid part of the self-energy, which provides only 
			small corrections above $\omega_D$, has been neglected for simplicity.
			L(R) stands for the left(right) branch of the outer band. 
			Index 1(2) corresponds to fits to $\Sigma_1$ ($\Sigma_2$).
			}
	\label{D_lamb}
\end{table}
%%%%%%%%%%%%%%%%%%%%%%%%%%%%%%%%

Fig.~\ref{fig:ZnO-2DEG-SigmaLEFT} shows the experimental dispersion 
and complex self-energy extracted from the left branch of the outer subband
of the 2DES in ZnO, together with fits using the 2D and 3D Debye models.
Table~\ref{D_lamb} present a summary of the fitting parameters 
for $\Sigma_1$ (index $1$) and $\Sigma_2$ (index $2$) obtained
from those two models for both the left (L) and right (R) branches 
of the outer subband of the 2DES in ZnO.

A comparison of Figs.~\ref{fig:ZnO-2DEG-Debye2D-3D} and~\ref{fig:ZnO-2DEG-SigmaLEFT},
and an inspection of the parameters listed in table\ref{D_lamb}, 
shows that the results are consistent with each other, 
except for a sensitively larger Debye frequency, and smaller coupling constant,
extracted from the fit to $\Sigma_2$ in the left branch of the outer subband ($L_2$ column). 
%%%
From all the other fits, 
the mean value of $\omega_D(2D)$ is $0.069$~eV, or $556.5$~cm$^{-1}$ in spectroscopy units, 
while the average of $\omega_D(3D)$ is $ 0.085$~eV or $685.6$~cm$^{-1}$.
The mean value for $\omega_D(2D)$ compares very well to the $E_1$ and $A_1$ LO modes 
identified in previous measurements of phonon modes in ZnO\cite{phonon_mode}. 
However, The mean value for $\omega_D(3D)$ does not correspond to any 
previously reported phonon energy in this material.
Thus, all in all, the 2D Fermi-liquid~$+$Debye model appears as a better description 
of our ARPES data on ZnO, coherent with previous results from other experimental probes.
%%%

%%%%%%%%%%%%%%%%%%%%%%%%%%%%%
\subsection*{2DES at the Zn-terminated ZnO surface}
%%%%%%%%%%%%%%%%%%%%%%%%%%%%%
%%%
%%%%%%%%%%%%%%%%%%%%%%%%%%%%%%%%
\begin{figure}[tb]
        \includegraphics[clip, width=0.48\textwidth]{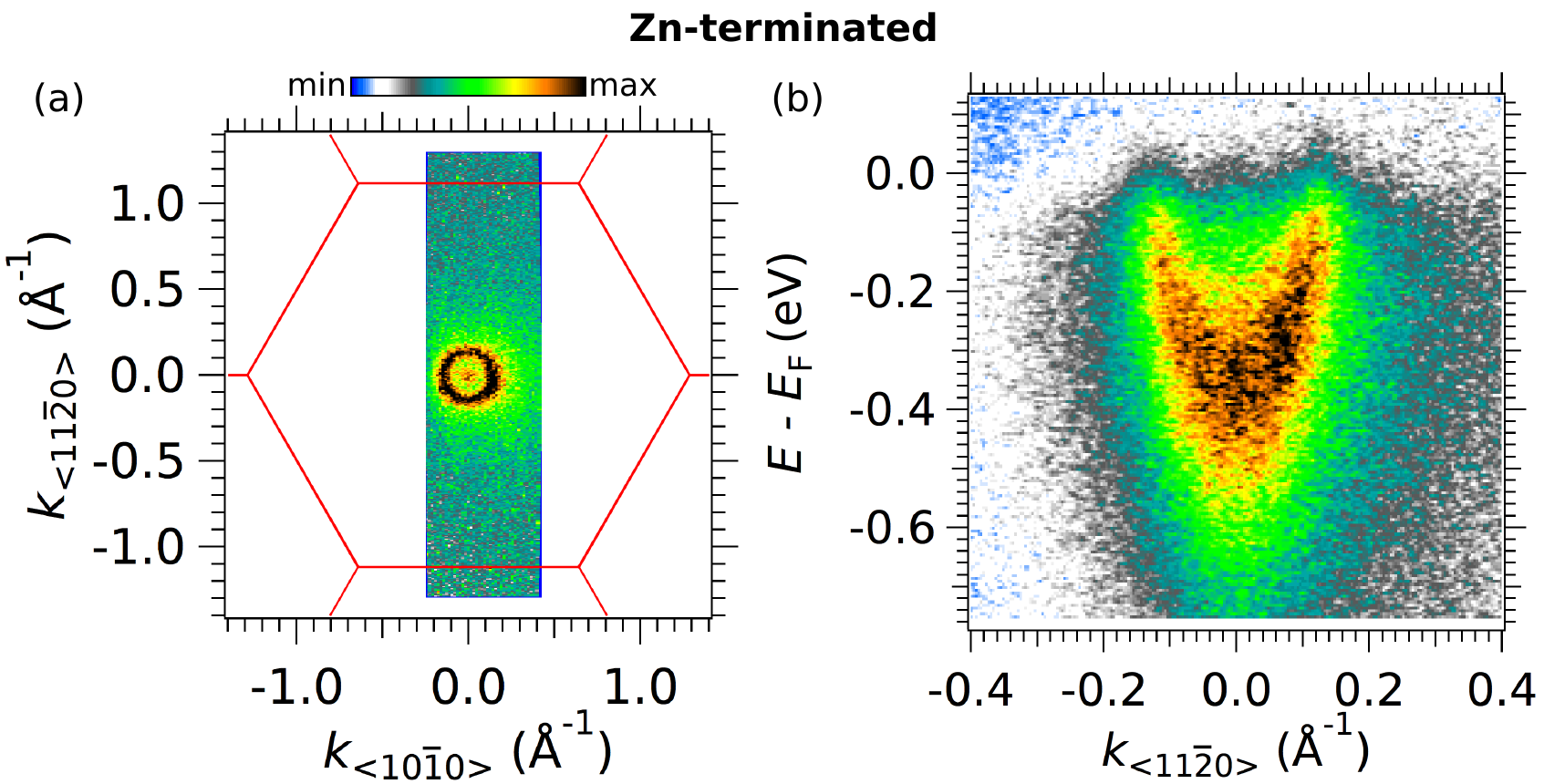}
    \caption{\label{fig:ZnO-2DEG-ZnTerm} {%(Color online)
        	 (a)~ARPES Fermi surface map of the Al(2~\AA)/ZnO interface
        	 in the Zn-terminated $[0001]$ plane.
        	 Red lines indicate the edges of the in-plane Brillouin zone.
			 (b)~Corresponding energy-momentum ARPES intensity map 
			 along the in-plane $k_{<1120>}$ direction.
			 All data in this figure were measured at $h\nu = 88$~eV, 
			 around the bulk $\Gamma_{003}$ point, 
			 with linear-horizontal light polarization
			 at $T = 7$~K. 
        }
      }
\end{figure}
%%%%%%%%%%%%%%%%%%%%%%%%%%%%%%%%
%%
Fig.~\ref{fig:ZnO-2DEG-ZnTerm}(a) shows the in-plane Fermi surface map measured by ARPES
at the Zn-terminated $[0001]$ plane of the Al(2~\AA)/ZnO$(0001)$ interface.
Similar to the O-terminated plane, this surface also shows two states
forming in-plane circular Fermi sheets around $\Gamma$.
However, their Fermi momenta are smaller than those obtained at the 
O-terminated surface.
Accordingly, as shown by the the energy-momentum dispersion map in 
Fig.~\ref{fig:ZnO-2DEG-ZnTerm}(b), the corresponding subbands disperse
down to smaller (in absolute value) energies. In particular, the bottom
of the inner subband is very close to $E_F$. 
%%

%%%%%%%%%%%%%%%%%%%%%%%%%%%%%%%%%%%%%%%%%%%%%%%%%%%%%%%%%%%%%%%%%%%%%%%%%%%%%%%%%%%%%%%
%%%%%%%%%%%%%%%%%%%%%%%%%%%%%%%%%%%%%%%%%%%%%%%%%%%%%%%%%%%%%%%%%%%%%%%%%%%%%%%%%%%%%%%
%%%%%%%%%%%%%%%%%%%%%%%%%%%%%%%%%%%%%%%%%%%%%%%%%%%%%%%%%%%%%%%%%%%%%%%%%%%%%%%%%%%%%%%

\end{document}